\documentclass[a4paper]{jpconf}
\usepackage{graphicx}

\usepackage{amsfonts}
\usepackage{amsmath}
\usepackage{amssymb}
\usepackage{graphicx}

\providecommand{\U}[1]{\protect\rule{.1in}{.1in}}
\newcommand{\ba}{\begin{eqnarray}}
\newcommand{\ea}{\end{eqnarray}}
\def\beq{\begin{equation}}
\def\eeq{\end{equation}}

\begin{document}

\begin{flushright}
CERN-TH-PH/2009-033\\
March 2009
\end{flushright}

\title{Exploration of Possible Quantum Gravity Effects with Neutrinos I:
Decoherence in Neutrino Oscillations Experiments}

\author{Alexander Sakharov$^{1,2,3}$, Nick Mavromatos$^{4}$,
Anselmo Meregaglia$^{5}$, Andr\' e Rubbia$^{2}$, Sarben Sarkar$^{4}$}

\address{$^{1}$ Theory Division, Physics Department, CERN, Geneva 23 CH 1211,
Switzerland\\
$^{2}$ Swiss Institute of Technology ETH-Z\"urich, 8093, Z\"urich,
Switzerland\\
$^{3}$ Department of Physics, Wayne State University, Detroit, MI 48202, USA\\
$^{4}$ Department of Physics, King's College London,
University of London, Strand, London, WC2R 2LS, United Kingdom \\
$^{5}$ IPHC, Universit, Louis Pasteur, CNRS/IN2P3, Strasbourg, France\\
}

\ead{Alexandre.Sakharov@cern.ch}

\begin{abstract}
Quantum gravity may involve models with stochastic fluctuations of the
associated metric field, around some fixed background value. Such stochastic
models of gravity may induce decoherence for matter propagating in
such fluctuating space time. In most cases, this leads to fewer
neutrinos of all active flavours being
detected in a long baseline experiment as compared to
three-flavour standard neutrino oscillations.
We discuss the potential of the CNGS and J-PARC beams in constraining models of
quantum-gravity induced decoherence using neutrino oscillations as a probe.
We use as much as possible model-independent parameterizations, even though they
are motivated by
specific microscopic models, for fits to the expected experimental data which
yield bounds on
quantum-gravity decoherence parameters.
\end{abstract}

\section{Introduction\label{sec:1}}

If microscopic black holes, or other defects forming space-time foam, exist
in the vacuum state of quantum gravity (QG)~\cite{wheeler,hawking}, this
state, in our view, will constitute an \textquotedblleft
environment\textquotedblright\ which will be characterised by some \textit{%
entanglement entropy}, due to its interaction with low-energy matter. 

The matter system in such a case behaves as an open quantum mechanical
system, exhibiting \emph{decoherence}, which has in principle detectable
experimental signatures. In the context of a phenomenological
parametrization of quantum-gravity induced decoherence the first tests along
these lines have been proposed in \cite{ehns} . A more microscopic
consideration was given in \cite{banks}, where the proposed parametrization
of decoherent effects of quantum gravity was forced to obey the Lindblad~%
\cite{lindblad,benattiinfstat,gorini} formalism of open systems, employing
completely
positive dynamical semigroup maps. This latter phenomenology, however, may
not be a true feature of a quantum theory of gravity.

In general, for phenomenological purposes, the important feature of such
situations is the fact that gravitational environments, arising from
space-time foam or some other, possibly semi-classical feature of QG, can
still be described by non-unitary evolutions of a density matrix $\rho $.
Such equations have the form $\partial _{t}\rho =\Lambda _{1}\rho +\Lambda
_{2}\rho$, where $\Lambda _{1}\rho =\frac{i}{\hbar }\left[ {\rho ,H}\right]$
and $H$ is the hamiltonian with a stochastic element in a classical metric.
Such effects may arise from back-reaction of matter within a quantum theory
of gravity \cite{ehns,hu} which decoheres the gravitational state to give a
stochastic ensemble description. Furthermore within models of D-particle
foam arguments~\cite{mavrik} in favour of a stochastic metric have been
given in~\cite{sarkar}. The Liouvillian term $\Lambda _{2}\rho $ gives rise to a
non-unitary
evolution. A common approach to $\Lambda _{2}\rho ,$ is to parametrise the
Liouvillian in a so called Lindblad form~\cite{lindblad,gorini} but this is
not based on microscopic physics. We note at this point that any non-linear
evolutions that may characterise a full theory of QG (see e.g. a
manifestation in Liouville strings~\cite{emnnl}), can be ignored to a first
approximation appropriate for the accuracy of contemporary experimental
probes of QG. Generically space-time foam and the back-reaction of matter on
the gravitational metric may be modelled as a randomly fluctuating
environment; formalisms for open quantum mechanical systems propagating in
such random media can thus be applied and lead to concrete experimental
predictions. The approach to these questions have to be phenomenological to
some degree since QG is not sufficiently developed at a non-perturbative
level.

One of the most sensitive probes of such stochastic quantum-gravity
phenomena are neutrinos~\cite{lisi,mavromatos,barenboim,barenboim2,
Benatti:2001fa,Benatti:2000ph,Brustein:2001ik,bmsw,lisib,lisi1}, and in
particular high-energy ones~\cite{winstanley,y}. For example,  as
pointed out recently in~\cite{barenboim}, the tiny mass differences between
neutrino flavours may themselves (in part) be the result of a CPT violating
quantum-gravity background. The phenomenon, if true, would be the
generalisation of the celebrated Mikheyev-Smirnov-Wolfenstein (MSW) effect~%
\cite{wolf,mikheev}. The latter arises from effective mass differences
between the various neutrino flavours, as a result of different type of
interactions of the various flavours with matter within the context of the
Standard Model. The phenomenon has been generalised to randomly fluctuating
media~\cite{loreti}, which are of relevance to solar and nuclear reactor $%
\beta $-decays neutrinos. This stochastic MSW effect will be more relevant
for us, since we consider space-time foam, as a random medium which induces
flavour-sensitive mass differences. If we can extrapolate~\cite{ms}
semi-classical results on black-hole evaporation, in both general relativity~%
\cite{gao} and string theory~\cite{lifschytz} to the quantum gravity foamy
ground state (assuming it exists and characterizes the ground state of some
(stochastic) quantum-gravity models, it follows that microscopic black holes
which are near extremal (and therefore electrically charged) would evaporate
significantly less, compared with their neutral counterparts. Thus, we may
assume~\cite{ms,barenboim}, that near extremal black holes in the foam would
\textquotedblleft live\textquotedblright\ longer, and as a result they would
have more time to interact with ordinary matter, such as neutrinos. Such
charged black holes would therefore constitute the dominant source of charge
fluctuations in the foam that could be responsible for foam-induced neutrino
mass differences according to the idea proposed in \cite{barenboim}. Indeed,
the emitted electrons from such black holes, which as stated above are
emitted preferentially, compared to muons or other charged particles, as
they are the lightest, would then have more time to interact (via coherent
standard model interactions) with the electron-neutrino currents, as opposed
to muon neutrinos. This would create a \textit{flavour bias} of the foam
medium, which could then be viewed~\cite{ms,barenboim} as the
\textquotedblleft quantum-gravitational analogue\textquotedblright\ of the
MSW effect~\cite{wolf,mikheev} in ordinary media (where, again, one has only
electrons, since the muons would decay quickly). In this sense, the quantum
gravity medium {can be} {partially responsible for generating
effective neutrino mass differences~\cite{barenboim}. As already indicated
by earlier phenomenological studies~\cite{bmsw} of quantum-gravity
induced decoherence models for neutrinos,  only a small part of the neutrino
mass differences and mixing can be attributed to interactions of the
neutrinos with the medium of the quantum-gravity space-time foam.
Nevertheless, the list of models examined so far~\cite{barenboim,ms,bmsw} is
not by any means an exhaustive list. Hence we consider the issue of the
effect of quantum gravity on the size of the  neutrino oscillation
parameters  an open one and worthy of further investigation. We also remark
that  in our quantum gravitational MSW scenario~\cite{barenboim,bmsw} the
charged black holes lead to a stochastically fluctuating medium.
Consequently we will adopt the formalism of the MSW effect for
stochastically fluctuating media~\cite{loreti}, where the density of
electrons is now replaced by the density of charged black hole/anti black
hole pairs. 

In what follows we report on study~\cite{deqrub} of decoherence induced by
non-linear space-time foam
fluctuations as a subdominant effect in neutrino oscillations at CNGS and
J-PARC beams after giving an overview of the framework of decoherence
phenomena in neutrino experiments. In particular in~\cite{deqrub} we
presented the damping signatures and the associated fitting
functions, which might be due to either the "quantum-gravitational analogue"
of the MSW effect or the stochastic fluctuations of the space-time metric
background. We considered various stochastic models of foam,
which lead to different damping signatures, depending on the details of the
underlying characteristic distribution functions~\cite{alexandre} and estimate
the sensitivity of CNGS and J-PARC
experiments to the parameters of quantum-gravitational decoherence entering
the set of the above-mentioned damping signatures.

\section{The combined fit to Quantum-Gravity Decoherence Signatures\label%
{sec:4}}

The various types of decoherence can be mainly distinguished by the form of
their exponential damping factor, as far as the power of the oscillation
length $L$ in the exponent is concerned, and the associated energy
dependence~\cite{dump}. Model independent data fits should combine, in
general, the various types of decoherence-deformed oscillations, given that
dominance of one or the other type may not be necessarily a feature of a
quantum-gravity model.

For our studies~\cite{deqrub} we use two sets of the one and two parametric
models
covering the main variety of phenomenologies for quantum gravity induced
decoherence phenomena. The
first set of the models under consideration concerns the presence of linear
Lindblad-type mapping operator in the equation for the evolution of the
density matrix for the pure neutrino quantum states \cite%
{lisi,lisib,Benatti:2001fa,Benatti:2000ph,ohlsson,mavromatos}. The
oscillation probabilities corrected for the decoherence effects with
different energy dependence in the exponentials read

\begin{itemize}
\item no neutrino-energy dependence 
\begin{equation}  \label{no_E}
P_{\nu_{\mu}\rightarrow\nu_{\tau}}=\frac{1}{2}\sin^{2}(2\theta_{23})\left[
1-\exp(-5\cdot10^{9}\gamma_{0}L) \cos\left( \frac{2.54\Delta m^{2}}{E}%
L\right) \right]
\end{equation}

\item inversely proportional to the neutrino energy (\emph{e.g.} the case of
Cauchy-Lorentz type of stochastic foam~\cite{alexandre}, (\ref{dampingCL})) 
\begin{equation}  \label{inv_E}
P_{\nu_{\mu}\rightarrow\nu_{\tau}}=\frac{1}{2}\sin^{2}(2\theta_{23})\left[
1-\exp(\frac{-2.54 \gamma_{-1}^{2}L}{E}) \cos\left( \frac{2.54\Delta m^{2}}{E%
}L\right) \right]
\end{equation}

\item proportional to the neutrino energy squared 
\begin{equation}  \label{sqr_E}
P_{\nu_{\mu}\rightarrow\nu_{\tau}}=\frac{1}{2}\sin^{2}(2\theta_{23})\left[
1-\exp(-5\cdot10^{27}\gamma_{2}E^{2}L) \cos\left( \frac{2.54\Delta m^{2}}{E}%
L\right) \right]
\end{equation}
\end{itemize}
where $\gamma _{0}$, $\gamma _{-1}^{2}$ and $\gamma _{2}$ are measured in
eV, eV$^{2}$ length and eV$^{-1}$ respectively the mass square difference $%
\Delta m^{2}$, is measured in eV$^{2}$, the energy $E$, is measured in GeV;
and the path, $L$, is measured in km.

The second set of the models concerns the gravitational MSW stochastic
effect~(\ref{2genprob}) with linear and quadratic time
dependent fluctuations of space-time foam described by 
\begin{equation}
P_{\nu _{\mu }\rightarrow \nu _{\tau }}=\frac{1}{2}-\exp (-\kappa _{1})\frac{%
\cos ^{2}(2\theta _{23})}{2}-\frac{1}{2}\exp (-\kappa _{2})\cos \left( \frac{%
2.54\Delta m^{2}}{E}L\right) \sin ^{2}(2\theta _{23}),  \label{msw}
\end{equation}%
where the exponential damping factors are chosen as

\begin{itemize}
\item no energy dependence, with linear 
\begin{equation}  \label{msw1}
\kappa_{1}=5\cdot10^{9}\alpha^{2}L\sin^{2} (2\theta);\ \kappa
_{2}=5\cdot10^{9}\alpha^{2}L(1+0.25(\cos(4\theta)-1))
\end{equation}

quadratic 
\begin{equation}  \label{msw2}
\kappa_{1}=2.5\cdot10^{19}\alpha_{1}^{2}L^{2}\sin^{2} (2\theta);\
\kappa_{2}=2.5\cdot10^{19}\alpha_{1}^{2}L^{2}(1+0.25(\cos (4\theta)-1))
\end{equation}

and combined time evolution 
\begin{align}  \label{msw1comb}
\kappa_{1}=(5\cdot10^{9}\gamma_{1}^{2}L+2.5\cdot10^{19}\gamma_{2}^{2}L^{2})%
\sin^{2} (2\theta);  \notag \\
\kappa_{2}=(5\cdot10^{9}\gamma_{1}^{2}L+2.5\cdot10^{19}%
\gamma_{2}^{2}L^{2})(1+0.25(\cos(4\theta)-1))
\end{align}

\item proportional to the neutrino energy, with linear 
\begin{equation}  \label{msw3}
\kappa_{1}=5\cdot10^{18}\beta^2 EL\sin^{2} (2\theta);\ \kappa
_{2}=5\cdot10^{18}\beta^2 EL(1+0.25(\cos(4\theta)-1))
\end{equation}

quadratic 
\begin{equation}  \label{msw4}
\kappa_{1}=2.5\cdot10^{28}\beta_{2}^{2} EL^{2}\sin^{2} (2\theta);\
\kappa_{2}=2.5\cdot10^{28}\beta_{2}^{2} EL^{2}(1+0.25(\cos (4\theta)-1))
\end{equation}

and combined time evolution 
\begin{align}  \label{msw2comb}
\kappa_{1}=(5\cdot10^{18}\gamma_{1}^{\prime2}EL+2.5\cdot
10^{28}\gamma_{2}^{\prime2}EL^{2})\sin^{2} (2\theta);  \notag \\
\kappa_{2}=(5\cdot10^{18}\gamma_{1}^{\prime2}EL+2.5\cdot10^{28}\gamma
_{2}^{\prime2}EL^{2})(1+0.25(\cos(4\theta)-1))
\end{align}

\item proportional to the neutrino energy squared, with linear time
evolution 
\begin{equation}  \label{msw5}
\kappa_{1}=5\cdot10^{27}\beta_{1}^{2} E^{2}L\sin^{2} (2\theta);\
\kappa_{2}=5\cdot10^{27}\beta_{1}^{2} E^{2}L(1+0.25(\cos (4\theta)-1))
\end{equation}
The energy and the path length in (\ref{msw1})-(\ref{msw5}) are measured in
GeV and km respectively, while the parameters in damping exponentials are
given in eV in respective power (see Table~\ref{table_cngs} for details).
\end{itemize}

\section{Sensitivity of CNGS and J-PARC beams to quantum-gravity decoherence 
\label{sec:cngs}}

In this section we study the expected sensitivity of the CNGS and J-PARC
beams to the quantum gravitational decoherence phenomena described by (\ref%
{no_E})-(\ref{msw5}), considering them as subdominant contributions to the
atmospheric oscillations effects.

Both CNGS and J-PARC are conventional neutrino beams where neutrinos are
produced by the decay of secondary particles (pions and kaons) obtained from
the collision of the primary proton on a graphite target. For the CNGS beam,
the protons come from the CERN-SPS facility with a momentum of 400 GeV/c
whereas in the case of the J-PARC~\cite{jparc} the protons are produced in
Tokay (Japan) and have a momentum of 40 GeV/c. The expected number of
protons on target per year at the nominal intensity is $4.5\times10^{19}$
and $1 \times10^{21}$ respectively for the CNGS and J-PARC beam and the
envisaged run length is 5 years in both cases.

Both beams will be used for long baseline neutrino experiments which,
starting from a $\nu_{\mu}$ beam, will search for neutrino oscillations. The
OPERA experiment will measure neutrino events on the CNGS beam using a 2
kton detector which relies on the photographic emulsion technique, located
at a baseline of 732 km; the first neutrino events were observed in August
2006~\cite{first_opera}.

The T2K experiment will use the J-PARC beam measuring neutrino events with
the Super-Kamiokande~\cite{sk_rev} detector (a water cerenkov detector with
an active volume of 22.5 kton) at a baseline of 295 km.

Although CNGS beam designed in a way to be optimised for the $\nu _{\mu
}\rightarrow \nu _{\tau }$ oscillation searches through the detection of $%
\tau $ lepton production in a pure $\nu _{\mu }$ beam there is also a
possibility to measure $\nu _{\mu }$ spectrum by reconstructing $\mu $ from
the charged current (CC) events caused by $\nu _{\mu }$. Moreover, for this
experiment, we can take advantage of high mean value for the energy of $\nu
_{\mu }$s which makes the exponential damping factors more pronounced for
some cases described in the previous section.

The number of $\mu$ is given by the convolution of the $\nu_{\mu}$ flux $%
d\phi_{\nu_{\mu}}/dE$ with the $\nu_{\mu}$ CC cross section on lead $%
\sigma_{\nu_{\mu}}^{\mathrm{CC}}(E)$, weigted by the $\nu_{\mu}\rightarrow%
\nu_{\mu}$ surviving probability $P_{\nu_{\mu}\rightarrow\nu_{\mu}}$, times
the efficiency $\epsilon_{\mu\mu}$ of muon reconstruction of a given
detector: 
\begin{equation}  \label{mu_count}
\frac{dN_{\mu\mu}}{dE}=A_{\mu\mu}\frac{d\phi_{\nu_{\mu}}}{dE}%
P_{\nu_{\mu}\rightarrow\nu_{\mu}}\sigma_{\nu_{\mu}}^{\mathrm{CC}}(E)
\epsilon_{\mu\mu}
\end{equation}
where $A_{\mu\mu}$ is a normalisation factor which takes into account the
target mass and the normalisation of the $\nu_{\mu}$ in physical units. In
our study we assumed an overall efficiency $\epsilon_{\mu\mu}$ of $93.5\%$
for the OPERA experiment and of $90\%$ for the T2K one as stated in the
experiment proposals.

To estimate quantitatively the sensitivity of CNGS on $P_{\nu _{\mu
}\rightarrow \nu _{\tau }}$ described by (\ref{no_E})-(\ref{msw5}), we
simulated the theoretical spectra of the reconstructed $\nu _{\mu }$ events
for various values of damping parameters (for details see~\cite{deqrub}). 
For the best-fit values~\cite{sk} of the atmospheric neutrino
parameters we used: $
\Delta m^{2}=2.5\cdot 10^{-3}\mathrm{eV^{2}}$; $\theta_{23}=45^{\circ }$.

 The 3~$\sigma $ sensitivity on the damping parameters is found by applying
a cut on the value of the $\chi ^{2}$ of 9 and 11.83 respectively for 1
d.o.f and 2 d.o.f.

As the CNGS beam is designed to observe $\nu _{\tau }$, neutrinos will have
a high energy with a mean value of about 17 GeV. This represents an
advantage since it makes the exponential damping factors more pronounced for
some cases described in the previous section. 

To generate the expected neutrino spectra of the CNGS beam measured by the
OPERA experiment we used a fast simulation algorithm described
in~\cite{sim_nu}. We present in Fig.\ref{numu_spectrum} a typical simulated
spectrum of the expected number of $\mu $
events including the effects of decoherence (for the case of an inversely
proportional dependence on neutrino energy) as a subdominant suppression of
the probability inferred from the atmospheric neutrino experiment~\cite{sk}.

\begin{figure}
\begin{minipage}{14pc}
\includegraphics[width=17pc]{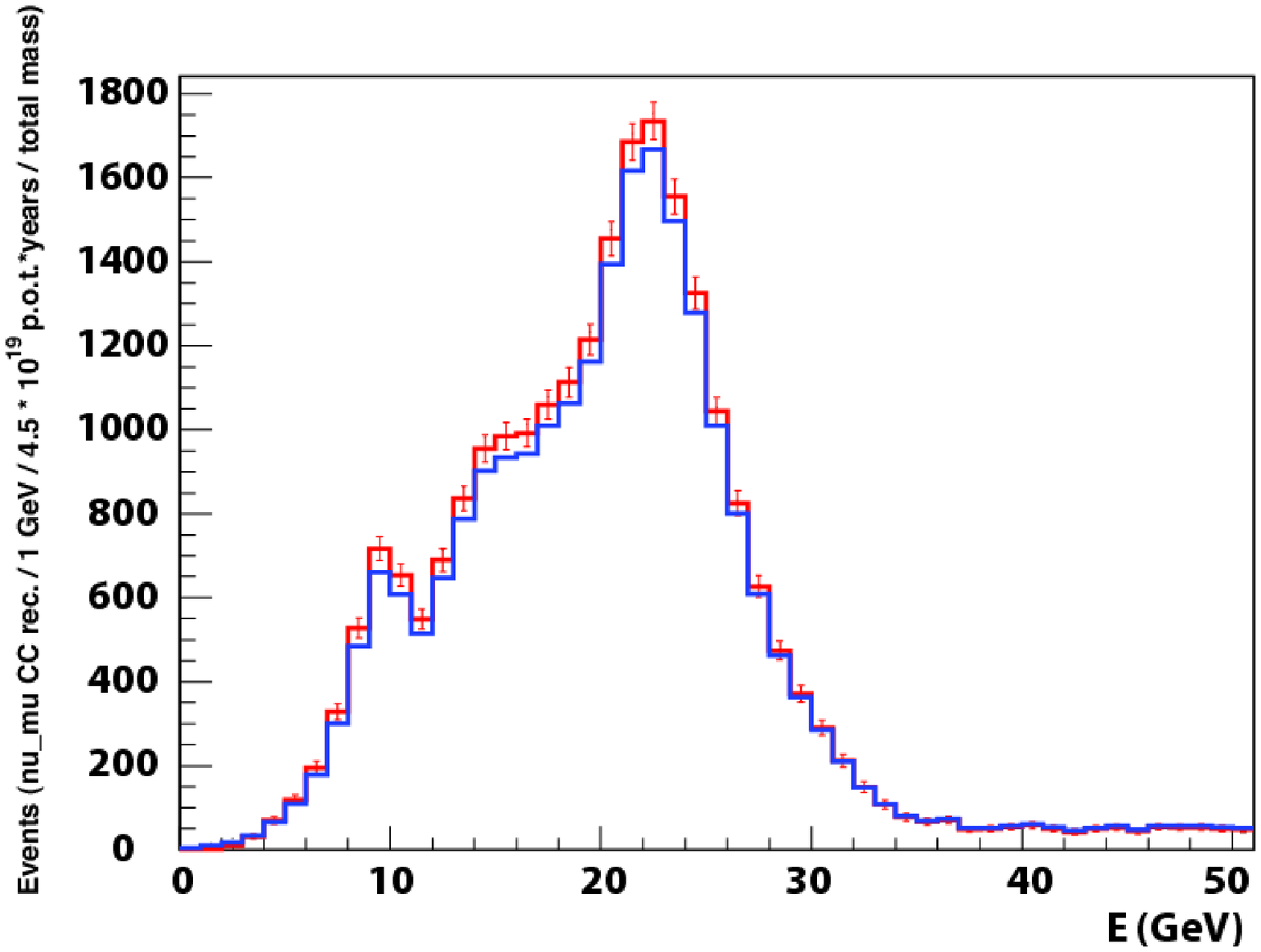}
\caption{\label{numu_spectrum}The number of reconstructed
$\protect\nu_{\protect\mu}$ CC events
in OPERA as a function of the neutrino energy with (blue line) and without
(red line with error bars) QG decoherence effect included in case of
inversely proportional dependence on neutrino energy. $3\protect\sigma$
difference between the expected and QG disturbed spectra is shown.}
\end{minipage}\hspace{2pc}%
\begin{minipage}{14pc}
\includegraphics[width=17pc]{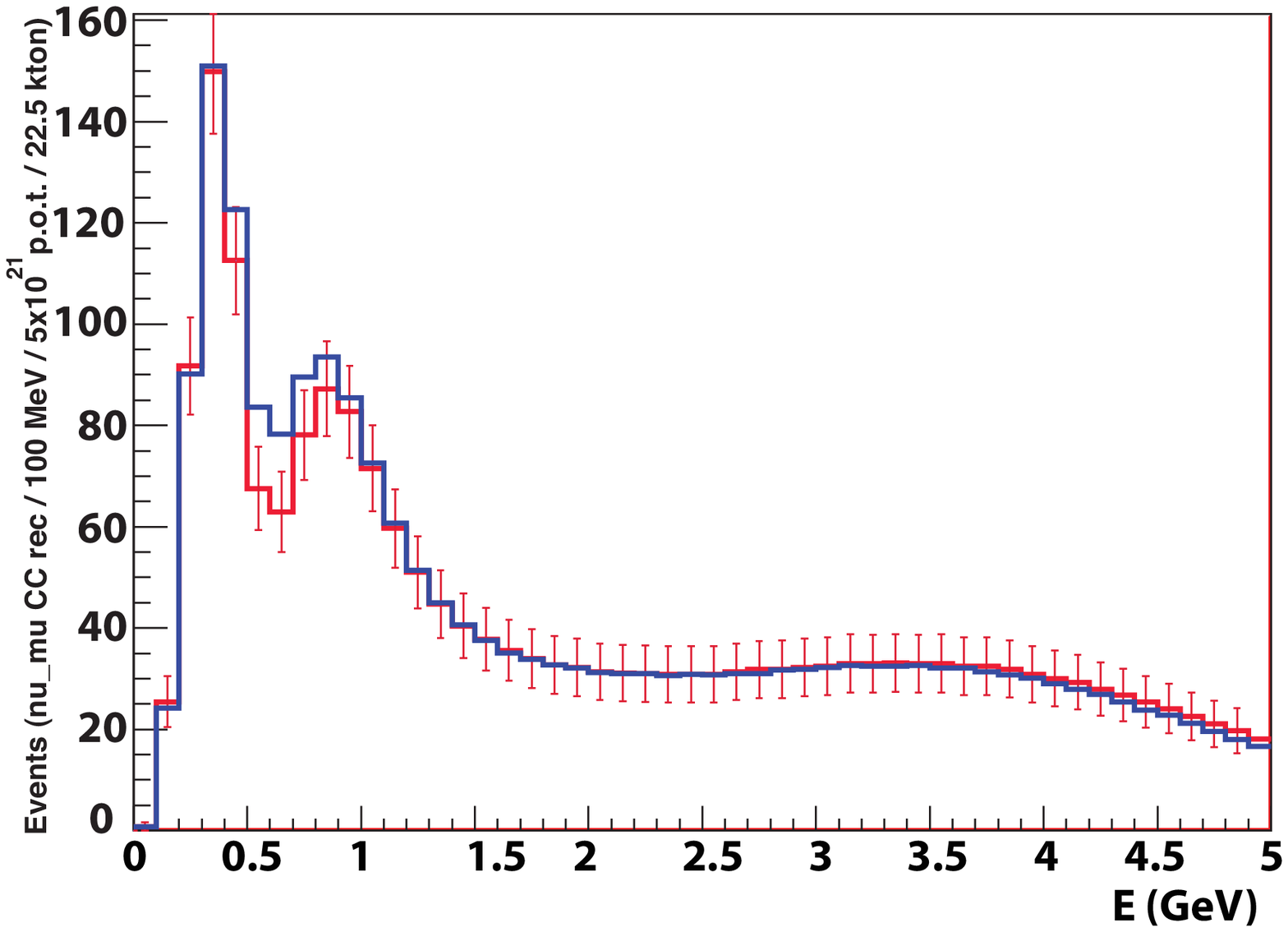}
\caption{\label{fig:T2K}The number of reconstructed $\protect\nu_{\protect\mu}$
CC events
in T2K as a function of the neutrino energy with (blue line) and without (red
line with error bars) QG decoherence effect included in case of inversely
proportional dependence on neutrino energy. $3\protect\sigma$ difference
between the expected and QG disturbed spectra is shown.}
\end{minipage} 
\end{figure}
\begin{figure}
\begin{minipage}{14pc}
\includegraphics[width=17pc]{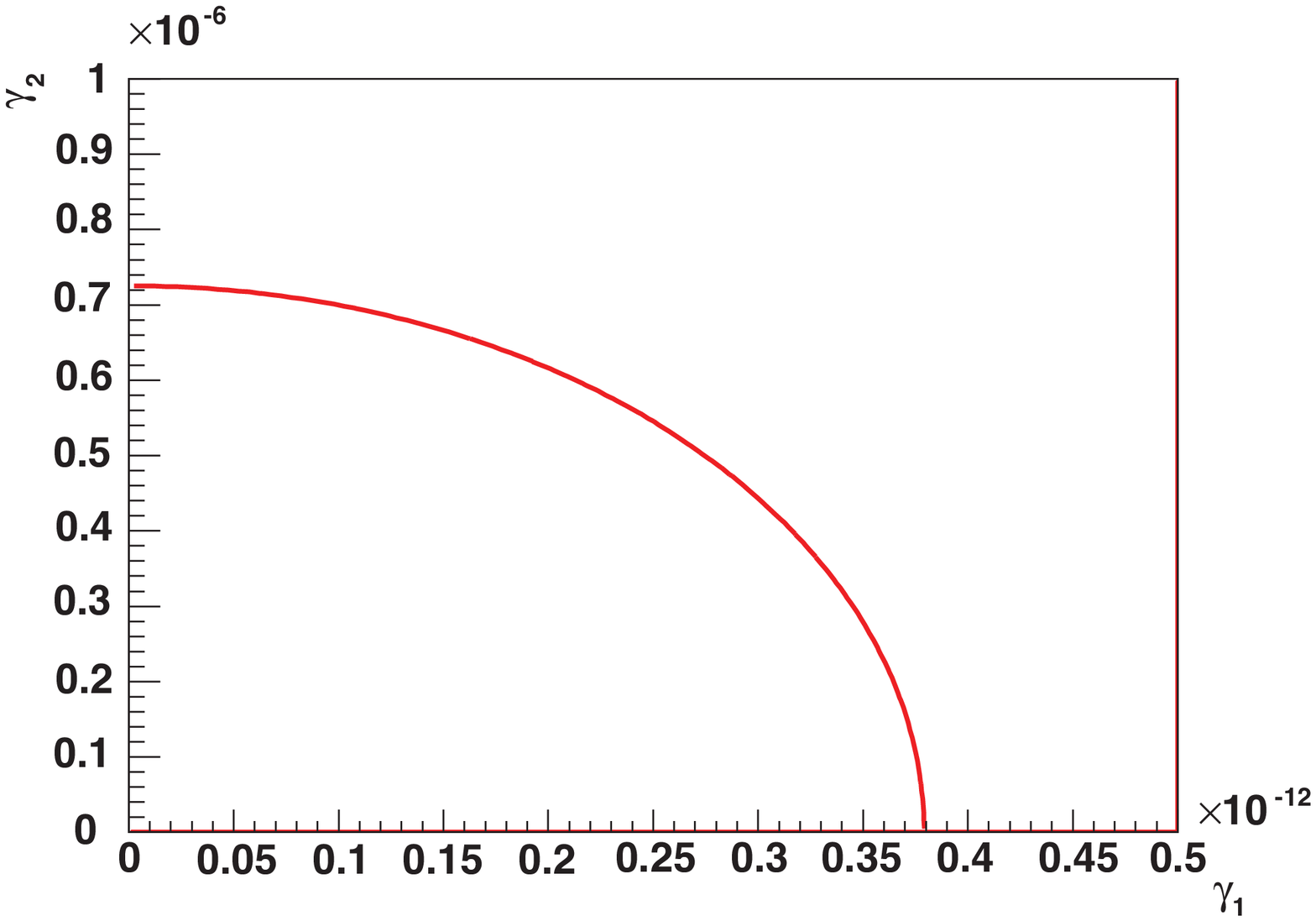}
\caption{\label{g1g2cngs_1} The expected CNGS sensitivity contour at
3~$\protect\sigma$ CL,
with two decoherence parameters contributing to the combined time evolution
of the gravitational MSW effect (with stochastic metric
fluctuations), calculated for damping with no energy dependence. }
\end{minipage}\hspace{2pc}%
\begin{minipage}{14pc}
\includegraphics[width=17pc]{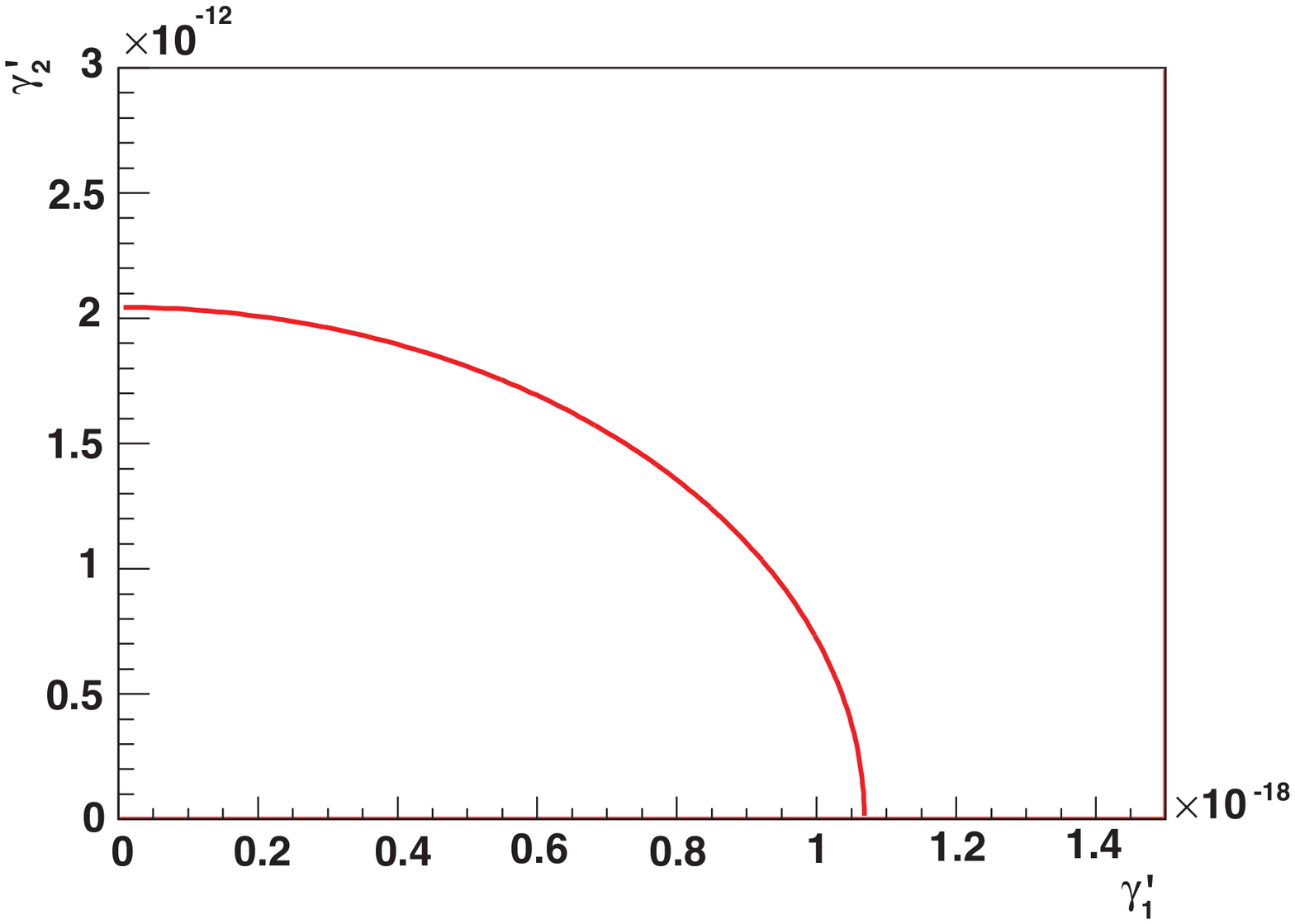}
\caption{\label{g1g2cngs_2} The expected CNGS sensitivity contour at
3~$\protect\sigma$ CL,
with two decoherence parameters contributing to the combined time evolution
of the gravitational MSW effect (with stochastic metric
fluctuations), calculated for damping proportional to the neutrino energy.}
\end{minipage} 
\end{figure}
{\ 
\begin{table}[t]
\begin{center}
{\scriptsize 
\begin{tabular}{|c|c|c|c|}
\hline
Lindblad-type mapping operators & CNGS & T2K & T2KK \\ \hline
\  &  &  & \  \\ 
$\gamma_{0}$\ $[\mathrm{eV}]$\ ;\ ($[\mathrm{GeV}]$) & $2\times10^{-13}$\ ;\
($2\times10^{-22}$) & $2.4\times10^{-14}$\ ;\ ($2.4\times10^{-23})$ & $%
1.7\times10^{-14}$\ ;\ ($1.7\times10^{-23}$) \\ 
\  &  &  & \  \\ 
$\gamma_{-1}^{2}$\ $[\mathrm{eV^{2}}]$\ ;\ ($[\mathrm{GeV^{2}}]$) & $%
9.7\times10^{-4}$\ ;\ ($9.7\times10^{-22}$) & $3.1\times10^{-5}$\ ;\ ($%
3.1\times10^{-23}$) & $6.5\times10^{-5}$\ ;\ ($6.5\times10^{-23}$) \\ 
\  &  &  & \  \\ 
$\gamma_{2}$\ $[\mathrm{eV^{-1}}]$\ ;\ ($[\mathrm{GeV^{-1}}]$) & $%
4.3\times10^{-35}$\ ;\ ($4.3\times10^{-26}$) & $1.7\times10^{-32}$\ ;\ ($%
1.7\times10^{-23}$) & $3.5\times10^{-33}$\ ;\ ($3.5\times10^{-24}$) \\ 
\  &  &  & \  \\ \hline
Gravitational MSW (stochastic) effects & CNGS & T2K & T2KK \\ \hline
\  &  &  & \  \\ 
$\alpha^{2}$ & $4.3\times10^{-13}~\mathrm{eV}$ & $4.6\times10^{-14}~\mathrm{%
eV}$ & $3.5\times10^{-14}~\mathrm{eV}$ \\ 
\  &  &  & \  \\ 
$\alpha_{1}^{2}$ & $1.1\times10^{-25}~\mathrm{eV^{2}}$ & $3.2\times10^{-26}~%
\mathrm{eV^{2}}$ & $6.7\times10^{-27}~\mathrm{eV^{2}}$ \\ 
\  &  &  & \  \\ 
$\beta^{2}$ & $3.6\times10^{-24}$ & $5.6\times10^{-23}$ & $1.7\times10^{-23}$
\\ 
\  &  &  & \  \\ 
$\beta_{2}^{2}$ & $9.8\times10^{-37}~\mathrm{eV}$ & $4\times10^{-35}~\mathrm{%
eV}$ & $3.1\times10^{-36}~\mathrm{eV}$ \\ 
\  &  &  & \  \\ 
$\beta_{1}^{2}$ & $8.8\times10^{-35}~\mathrm{eV^{-1}}$ & $3.5\times10^{-32}~%
\mathrm{eV^{-1}}$ & $7.2\times10^{-33}~\mathrm{eV^{-1}}$ \\ 
\  &  &  & \  \\ \hline
\end{tabular}
}
\par
{\scriptsize \vspace{0.3cm} }
\end{center}
\caption{Expected sensitivity limits at CNGS, T2K and T2KK to one parametric
neutrino decoherence for Lindblad type and gravitational MSW (stochastic
metric fluctuation) like operators~\cite{deqrub}. {These results are obtained
for
the ``true'' values of the oscillation parameters fixed at $\Delta
m^{2}=2.5\cdot 10^{-3}\mathrm{eV^{2}}$ and $\protect\theta_{23}=45^{\circ }$~%
\protect\cite{sk}}.}
\label{table_cngs}
\end{table}
}
Contrary to the OPERA experiment, the T2K experiment was designed to observe 
$\nu _{e}$ and the mean energy is much lower: the maximum of oscillation at
the given baseline of 295 km corresponds to a neutrino energy of about 600
MeV and a narrow spectra at the selected energy will be obtained using the
so called off-axis technique~\cite{off-axis}. The spectrum covers the region
of the first maximum of oscillation and this is a region where the QG
effects could be easily observed due to the small number of $\nu _{\mu }$ CC
events expected in case of no QG damping exponents, as it can be seen in Fig~%
\ref{fig:T2K}.

Our results for the sensitivity of CNGS to one parametric decoherence
damping exponentials in $P_{\nu _{\mu }\rightarrow \nu _{\tau }}$ are
summarised in second column of Table~\ref{table_cngs}. Also, for two
parametric fits~(\ref{msw1comb}) and~(\ref{msw2comb}), the 3~$\sigma $ CL
sensitivity contours are presented in Fig.~\ref{g1g2cngs_1} and
Fig.~\ref{g1g2cngs_2}.

Our results obtained using the same way of analysis quantified
by~(\ref{mu_count}) for the sensitivity of T2K to one parametric
decoherence damping exponentials in $P_{\nu_{\mu}\rightarrow\nu_{\tau}}$ are
summarised in third column of Table~\ref{table_cngs}. Also, for two
parametric fits~(\ref{msw1comb}) and~(\ref{msw2comb}), the 3~$\sigma$ CL
sensitivity contours are similar to those obtained for CNGS.

The T2K experiment yields a better limit on the damping parameters only in
the case where the effect has no energy dependence or contains inversely
proportional to the neutrino energy exponent, as expected given the low
energy spectrum. In all the other cases, the dependence on the baseline
disfavours the short baseline of T2K with respect to OPERA.

Another possibility to observe the effect on the T2K neutrino beam is to
select a longer baseline, namely to locate the detector at about 1000 km in
Korea. Studies of beam upgrades and a large liquid Argon detector of 100
kton in Korea were carried out~\cite{GlacierKorea} in the framework of CP
violation discovery. We considered this option, called T2KK, and studied the
possibility to constrain damping parameters in this case. The proposed
upgrade at 4 MW of the beam was taken into account which results into $%
7\times 10^{21}$ p.o.t. per year and a running time of 4 years was
envisaged. 

Our results for the sensitivity of T2KK to one parametric decoherence
damping exponentials in $P_{\nu _{\mu }\rightarrow \nu _{\tau }}$ are
summarised in fourth column of Table~\ref{table_cngs}. This configuration
yields better results than the T2K experiment and results comparable to the
OPERA experiment. {All bounds obtained in Table~\ref{table_cngs} are
evaluated at the best-fit oscillation parameters given by~(\ref{atm_true}).

%%%%%%%%%%%%%%%%%%%%%%%%%%%%%%%%%%

\section{Conclusions}

It is instructive to compare the sensitivity limits presented in Table~\ref%
{table_cngs} with those derived from the analysis of atmospheric neutrino
data~\cite{lisi} obtained at Super-Kamiokande and K2K experiments. The numbers
of Table~\ref{table_cngs} in parentheses can be directly
compared with the bounds~\cite{lisi}.  In particular,
the bound obtained in~\cite{lisi} at 95\%
C.L. on the Lindblad type operators with no energy dependence is close to
the sensitivity estimated in our analysis in case of T2K and T2KK
simulations. Although, the CNGS estimation is about an order of magnitude
weaker, one should stress that the current limit is given at 99\% C.L. under
the assumption of the most conservative level of the uncertainty of the
overall neutrino flux at the source. The bound on the inverse energy
dependence given in~\cite{lisi} is close to the current CNGS
estimates. T2K and T2KK demonstrate an improvement. In spite of the fact
that the Super-Kamiokande data contains neutrino of energies up to $\sim$%
TeV, the sensitivity one obtains at CNGS to the energy-squared dependent
decoherence is close, within an order of magnitude, to the bound
) imposed by atmospheric neutrinos and surpasses T2K and T2KK sensitivity
bounds by $\approx 3$ and $\approx 2$ orders of magnitude respectively. The
much less uncertain systematics of CNGS compared to the atmospheric neutrino
data will make the expected bound more robust as soon as the upcoming data
from OPERA will be analysed. Moreover, our results are also competitive with
the sensitivity to the same Lindbland operators estimated in~\cite{morgan}
for ANTARES neutrino telescope, which is supposed to operate at neutrino
energies much higher than CNGS and J-PARC experiments.

Assuming that the decoherence phenomena affect all particles in the same
way, which however is by no means certain, one might compare the results of
our analysis with bounds obtained using the neutral kaon system~\cite{cplear}%
. The comparison could be done for the constant (no-energy dependence)
Lindblad decoherence model. The main bound in~\cite{cplear} in such a case
reads $\gamma_{0}\le4.1\times10^{-12}$~eV, thus being about two orders of
magnitude weaker than the sensitivity forecasted in the present paper.

Finally, we compare the estimated sensitivity with the bounds obtained in~%
\cite{lisi1} using solar+KamLAND data. In principle, as in the case of the
neutral kaon system, a direct comparison is impossible, since the parameters
investigated here for the $\nu_{\mu}\rightarrow\nu_{\tau}$ channel need not
be the same for the $\nu_{e}\rightarrow\nu_{\mu}$ channel. However, again,
if these parameters are assumed to be roughly of equal size, then one can
see that the estimates of~\cite{lisi1}, which win essentially
over the CNGS, T2K and T2KK sensitivities only for the case of inverse
energy dependent decoherence, which strongly favours low neutrino energies (%
\emph{e.g.} the case of Cauchy-Lorentz stochastic space-time foam models
of~\cite{alexandre}, for which the current limit would bound,
on account of, the scale parameter $\xi $ of the distribution to (for details
see \cite{alexandre,deqrub}): $\xi < 5 \times 10^{-3}$ for neutrino-mass
differences~%
\cite{lisi1} $|\mathrm{m}^2_{\mathrm{e}} - \mathrm{m}^2_\mu| = (7.92 \pm
0.71 )\times 10^{-5}$ eV$^{2}$). For the completeness, we mention that, our
best expected bound on the inverse-energy decoherence will imply, the
bound on the $\nu_{\mu}$ life time $%
\tau_{\nu_{\mu}}/m_{\nu_{\mu}}>3\times 10^{22}$~$\mathrm{GeV}^{-2}$.

The precise energy and length dependence of the damping factors is an
essential step in order to determine the microscopic origin of the induced
decoherence and disentangle genuine new physics effects from conventional
effects, which may also contribute
to decoherence-like damping. Some genuine quantum-gravity effects, such as
MSW like effect induced by stochastic fluctuations of the space-time, are
expected to increase in general with the energy of the probe, as a result of
back reaction effected on space-time geometry, in contrast to
ordinary-matter-induced `fake' CPT violation and `decoherence-looking'
effects, which decrease with the energy of the probe~\cite{dump}. At
present the sensitivity of the
experiments is not sufficient to unambiguously determine the microscopic
origin of the decoherence effects, but according to our estimations of the
most plausible energy-length dependencies for the MSW like decoherence the
sensitivity of CNGS and T2K will improve the current limits by at least two
orders of magnitude and one would arrive at definite conclusions on this
important issue. Thus phenomenological analyses like ours are of value and
should be actively pursued when the data from OPERA and T2K will become
available.

\ack
Alexander Sakharov expresses his gratitude to the organizers of
DICRETE'08 for financial support, warm hospitality and enthusiastic
scientific atmosphere of the meeting.

\end{document}